\documentclass[12pt]{article}

\usepackage{amsmath}
\usepackage{amssymb}
\usepackage{latexsym}
\usepackage{graphics}
\usepackage{psfrag,fancyhdr,epsfig}

\begin{document}

\begin{titlepage}

\vspace*{1cm}
\begin{center}
{\bf \Large Pseudo-3-Branes in a Curved 6D Bulk}

\bigskip \bigskip \medskip

{\bf C. Bogdanos}$^1$, {\bf A. Kehagias}$^2$ and {\bf K. Tamvakis}$^1$

\bigskip
$^1$ {\it Department of Physics,\\
University of Ioannina, Ioannina GR-45110, Greece}

$^2$ {\it Physics Division,\\
National Technical University of Athens, Zografou GR15780, Greece}

\bigskip \medskip
{\bf Abstract}
\end{center}
We consider a model involving  a 4-brane in a 6D bulk which
carries sigma model fields. An axion field on the 4-brane cancels
the pressure along one direction leading to an effective
codimension-2 3-brane. For a range
 of parameters of the theory, we get a transverse space which is non-compact,
 providing a possible solution to the cosmological constant problem. A setup
with two branes in a
  compact space is also treated. In this case, a mild fine-tuning between the radii of the
   two 4-branes is necessary. Finally, we explore the 4-brane problem in the Gauss-Codazzi formulation and
   we discuss general aspects of gravity in the presence of additional brane sources.

\end{titlepage}

\section{Introduction}

The issue of the cosmological constant appears
 to be one of the most pressing problems in modern theoretical physics, permeating many
  areas, ranging from field theory to cosmology. Conventional
   estimates of its value based on field theoretical approaches are
    in gross contradiction with observations (around 120 orders of
magnitude higher than observed for a Planck-scale cutoff). This
apparent failure has led to a variety of attempts to attack
     the problem that have yielded a number of new and exciting ideas
      in both basic theory and model building. The emergence of brane models and theories of large extra
dimensions
\cite{Randall:1999vf},\cite{Randall:1999ee},\cite{ArkaniHamed:1998rs}
in recent years has been a major driving force in gravitational
physics and a source of rich and interesting phenomenological
studies. One of the most interesting feature of these theories is
that they provide a possible explanation for the smallness of the
cosmological constant~\cite{Dvali:2002pe}. Similarly, the simple
observation that a codimension 2 brane is always Ricci-flat, opend
another possibility for the cancellation of the cosmological
constant. As it was shown in \cite{Carroll}, a four-dimensional
brane embedded in a six-dimensional bulk space, where the
transverse space is compactified using fluxes, induces a conical
singularity in the bulk, deforming the space into a rugby-ball
configuration. In this way, the brane cosmological constant
(tension) is completely offloaded into the bulk, where it
determines the deficit angle of the induced conical singularity.
Although this setup only partially solves the problem, as
fine-tunings are in the end needed to ensure a flat brane
\cite{Csaki:2000wz},\cite{Vinet:2004bk},\cite{Navarro:2003vw}, it
nevertheless provides evidence that extra dimensions may play a
role in resolving the cosmological constant conundrum. Extensions
and alternatives were consequently investigated by several authors
(see \cite{Navarro:2003vw}-\cite{Navarro:2003bf} and references
therein).

In this paper we study a model of an effective codimension-2
3-brane  embedded in a 6-dimensional bulk space. It is a
pseudo-3-brane in the sense that although  the submanifold within
which ordinary matter and fields are confined has in fact 5
dimension, i.e. is a 4-brane, the energy-momentum tensor of these
fields is selected in such a way to resemble the tensor structure
of a 3-brane.
 The presence of the one additional dimension of the brane is in this
  way hidden from the exterior space and we can take advantage of the well
   known result that branes of codimension-2 only produce conical singularities
    in their bulk spaces \cite{Deser:1983tn}-\cite{Klebanov:2007us}.
    In order to produce the required energy momentum tensor we will employ
    the technique discussed in \cite{Kaloper:2007my},\cite{Kaloper:2007ap}.
    However, we will not assume an empty bulk space, but one endowed with an $O(3)$
    sigma model (see  \cite{Kehagias:2004fb},\cite{RandjbarDaemi:2004ni},\cite{GellMann:1984mu}), unlike the usual flux compactification using $U(1)$
    fields \cite{Salam:1984cj},\cite{Sundrum:1998ns}. The transverse space away from the brane will
     thus be curved, but its curvature will only be due to the sigma model fields, since the brane
      cannot contribute more than a conical singularity at its position. This is a setup similar to
       the one discussed in \cite{Kehagias:2004fb}, the main difference being the existence of
       a resolved brane, rather than a purely codimension-2 singularity.

  The purpose of such a setup is to address the cosmological constant problem.
  We will see that, by having an appropriate energy content, we can construct
  an effectively flat 4-geometry on the brane, with all curvature being pushed
  away into the transverse space. To avoid imposing unsettling fine-tunings, we
  will allow for a space which is non-compact. We thus have one dimension tangent
  to the 4-brane which is not compact but can be integrated to a finite volume and
  a parallel dimension which is compactified in the spirit of Kaluza-Klein theories.
  As it turns out, for a range of parameters of the model, the 2-dimensional transverse
  space is normalizable and we thus expect a conventional KK phenomenology, with a zero
  mode mediating ordinary four-dimensional gravity on the brane and a spectrum of KK
  excitations as corrections. The cosmological constant of the 3-brane can be set to
  zero by proper choice of parameters, which induces a corresponding deficit angle in
  the bulk. Once this choice is made, any variation in one of the model's parameters,
  e.g. the brane cosmological constant, results to a new value for the deficit angle,
  without any additional fine-tuning. In the last section of our paper, we derive
   Einstein's equations on the 4-brane by employing the Gauss-Codazzi formalism.
   Based on these equations we discuss general aspects of such models and also the
   effects of additional brane matter, which will be the source for conventional 4-D
   gravity once the KK reduction of the compact dimension is performed.

\section{General Setup}
  We will start by describing the action of the setup. It includes a non-linear sigma model targeted
  on a K\"ahler manifold inside a six-dimensional bulk space. The bulk cosmological constant is assumed
  to be zero. A 4-brane is also included, carrying a tension $\sigma$ and an axion field $\Sigma$. The
axion is used to counteract the brane tension along the azimuthal
direction $\varphi$ and to ensure that the energy-momentum tensor
of the 4-brane mimics a 3-brane outside the resolved brane core,
so that we get an effective codimension-2
brane\cite{Kaloper:2007ap}. The metric of the bulk space is
considered to be
\begin{equation}
ds_6 ^2  = n_{\mu \nu } dx^\mu  dx^\nu   + d\rho ^2  + g_{\varphi
\varphi }(\rho) d\varphi ^2
\end{equation}
in  Gauss-normal coordinates. Note that only $\rho$ is the
direction transverse to the 4-brane, while $\varphi$ is our
compact dimension. The brane is situated at a distance $\rho_0$
away from the origin of the transverse space
 in the radial direction. The action of the model is thus of the form
\[
S = \int {d^6 x\sqrt { - g} \left( {M^4 R - \frac{1}{{2\lambda ^2
}}h_{\alpha \beta } \left( \phi  \right)\nabla ^M \phi^\alpha
\nabla _M \phi ^\beta  } \right)}
\]
\begin{equation}
 - \int {d^5 xd\rho \delta \left( {\rho  - \rho _0 } \right)\sqrt { - \gamma }
  \left( {\sigma  + \frac{1}{2}\gamma ^{\alpha \beta } \partial _\alpha  \Sigma \partial _\beta  \Sigma } \right)}\,\,.
\end{equation}
Uppercase Greek indices run from 0 to 5, while lowercase are
indices on the 4-brane that do not include the $\rho$ coordinate
and $\alpha, \beta$ are indices in the K\"ahler manifold. The
first term represents the bulk contributions from the
gravitational sector and the sigma model fields $\phi^{\alpha}$,
while the second contains the brane contributions for the tension
and the axion field. For the moment we assume a zero energy
content from ordinary matter on the brane.  The energy-momentum
tensor of the brane can be written as
\begin{equation}
T^{\left( {br} \right)} _{MN}  = \delta \left( {\rho - \rho_0 }
\right)\frac{{\sqrt { - \gamma } }}{{\sqrt { - g} }}\delta ^\mu  _M \delta ^\nu  _N
\left( { - \sigma \gamma _{\mu \nu }  - \frac{1}{2}\gamma _{\mu \nu }
\gamma ^{\alpha \beta } \partial _\alpha  \Sigma \partial _\beta  \Sigma  +
 \partial _\mu  \Sigma \partial _\nu  \Sigma } \right)\,.
\end{equation}
We also have an energy momentum tensor for the $\phi^{\alpha}$ fields
\begin{equation}
T^{\left( \phi  \right)} _{MN}  = \frac{{h_{\alpha \beta } }}{{\lambda ^2 }}
\left( {\nabla _M \phi ^a \nabla _N \phi ^\beta   -
 \frac{1}{2}g_{MN} \nabla ^\Lambda  \phi ^a \nabla _\Lambda  \phi ^\beta  } \right)\,.
\end{equation}
Einstein's equations become in this case
\begin{equation}
R_{MN}  - \frac{1}{2}g_{MN} R = \frac{1}{{2M^4 }}\left( {T^{\left( \phi  \right)} _{MN}  + T^{\left( {br} \right)} _{MN} } \right)
\end{equation}
or
\begin{equation}
R_{MN}  = \frac{1}{{2M^4 }}\left( {T^{\left( \phi  \right)} _{MN}  -
 \frac{1}{{N - 2}}T^{\left( \phi  \right)}g_{MN} } \right) + \frac{1}{{2M^4 }}
 \left( {T^{\left( {br} \right)} _{MN}  - \frac{1}{{N - 2}}T^{\left( {br} \right)}g_{MN} } \right)\,,
\end{equation}
where we denote by $N$ the total bulk dimension and we use $n$ for
the dimensions of the brane (in the case we will examine, $N=6$
and $n=5$). The energy-momentum tensor contains contributions from
both the complex scalar and the brane content. The axion field
equations are solved by  $$\Sigma  = q\varphi$$ so that $\Sigma$
has  $2\pi q$ jumps as we go around the $\varphi$ direction. If in
addition, the parameter $q$ (the axion charge) is such that
\begin{equation}
q^2= 2 \sigma g_{\varphi\varphi} \, , \label{tuning1}
\end{equation}
the axion contribution completely eliminates any tension along the
azimuthal direction $\varphi$. Notice that the above condition is
not a fine-tuning between $q$ and $\sigma$ in the usual sense,
since the metric component carries integration factors
 which are not determined yet by other constraints. Thus, the value of $q$ is not fixed against the brane
  tension. With this choice we get,
\[
T^{\left( {br} \right)} _{MN}  - \frac{1}{{N - 2}}g_{MN} T^{\left(
{br} \right)}  = \delta \left( {\rho  - \rho _0}
\right)\frac{{\sqrt { - \gamma } }}{{\sqrt { - g} }}\left[ {\sigma
\left( { - \gamma _{\mu \nu } \delta ^\mu  _M \delta ^\nu  _N  +
\frac{n}{{N - 2}}g_{MN} } \right)} \right.
\]
\begin{equation}
\left. { + \frac{1}{2}\gamma ^{\alpha \beta } \partial _\alpha
\Sigma \partial _\beta  \Sigma \left({ - \gamma _{\mu \nu } \delta
^\mu  _M \delta ^\nu  _N  + \frac{{n - 2}}{{N - 2}}g_{MN} }
\right) + \delta ^\mu  _M \delta ^\nu  _N \partial _\mu  \Sigma
\partial _\nu  \Sigma } \right]
\end{equation}
and similarly for the scalar fields
\begin{equation}
T^{\left( \phi  \right)} _{MN}  - \frac{1}{{N - 2}}g_{MN} T^{\left( \phi  \right)}  =
\frac{{h_{\alpha \beta } }}{{\lambda ^2 }}\nabla _M \phi ^\alpha  \nabla _N \phi ^\beta\,\,.
\end{equation}
Since we are using Gauss-normal coordinates, the determinants of
the bulk and the induced metric are the same and their ratio
cancels. We also have the equation of motion for the scalar
fields, which reads
\begin{equation}
\nabla ^M \nabla _M \phi ^\alpha   +
 \Gamma ^\alpha  _{\beta \gamma } \nabla ^\Gamma  \phi ^\beta  \nabla _\Gamma  \phi ^\gamma   = 0\,\,.
\end{equation}

\section{Single and Double Brane Solution}
In order to proceed in solving the equations in the bulk, we must
choose an ansatz for the bulk fields and the transverse metric. We
will assume an ${\textit O}(3)$ sigma model and a metric for the
K\"ahler manifold, which is now an ${\textit S^{ 2}}$ sphere, of
the form
\begin{equation}
h_{\alpha \beta }  = \frac{4}{{\left( {1 + \phi ^2 } \right)^2
}}\delta _{\alpha \beta }
\end{equation}
where $\phi ^2  = \left( {\phi ^1 } \right)^2  + \left( {\phi ^2 }
\right)^2$. This particular choice of the sigma model is entirely
{it ad hoc} and has been taken as a simple example. Another choice
of the sigma model target space of the form appearing in
supergravity like $SL(2)/U(1)$  would be equally good. In order to
investigate the resulting geometry of the transverse 2-D space, we
will reparametrize it in a conformally flat fashion as
\begin{equation}
ds_2 ^2  = d\rho ^2  + g_{\phi \phi } d\varphi ^2  = \psi \left( r \right)\left( {dr^2  + r^2 d\varphi^2 } \right) =
\psi \left( r \right) \delta_{mn} dy^m dy^n \,\,.
\label{coords}
\end{equation}
This change of coordinates implies that $d\rho ^2  = \psi \left( r
\right)dr^2$, which leads to
\[
 \Rightarrow \delta \left( {\rho  - \rho _0 } \right) = \frac{1}{{\sqrt
 {\psi \left( r \right)}}}\delta \left( {r - r_0 } \right)
\]
 for the transformation of delta function, where $r_0$ is the
position corresponding to $\rho_0$ in the new conformal
coordinates.  We will now adopt an ansatz for the sigma model
fields of the form $\phi ^\alpha   = y^\alpha$, such that
$\phi^2=r^2$. This ansantz solves the equations for the scalar
fields without any further constraints and from the expressions
for the scalar energy-momentum tensor we see that only the
$T^{(\phi)}_{\mu \nu}$ components survive. Einstein's equations
reduce to
\begin{equation}
R_{\mu \nu }  = 0\,,
\end{equation}
\begin{equation}
R_{mn}  = \frac{2}{{M^4 \lambda ^2 \left( {1 + r^2 } \right)^2 }}\delta _{\alpha \beta } \nabla _m \phi ^\alpha  \nabla _n
\phi ^\beta  \, + \,\frac{{2\sigma }}{{M^4 }}\frac{{g^{\left( 2 \right)} _{mn} }}{{\sqrt {\psi \left( r \right)} }}\,
\delta \left( {r - r _0 } \right)\,,
\end{equation}
where $\mu,\nu=0,1,2,3$ are coordinates on the effective 3-brane
and $m,n$ are coordinates of the 2-dimensional transverse space.
Note again that only $r$ is truly transverse to the brane. The
first set of equations ensures a flat 4-dimensional space. The
transverse 2-space will be curved. Taking the trace of the last
equation yields
\begin{equation}
R^{\left( 2 \right)}  = \frac{4}{{M^4 \lambda ^2 }}\frac{1}{{\psi \left( r \right)\left( {1 + r^2 } \right)^2 }} \,+ \,
\frac{{4\sigma }}{{M^4 }}\frac{1}{{\sqrt {\psi \left( r \right)} }}\,\delta \left( {r - r _0 } \right)\,,
\label{ricci2}
\end{equation}
where $R^{\left( 2 \right)}  =  - \frac{1}{\psi }\nabla ^2 \ln \psi $ is the Ricci scalar of the 2-D transverse space.
 Since the conformal factor $\psi (r)$ depends only on the radial coordinate, when it multiplies the
delta function, it becomes just a constant $\psi(r_0)$. The solution in the absence of the brane is known to be
\begin{equation}
\psi \left( r \right) = C_2 \frac{{r^{\frac{2}{{M^4 \lambda ^2 }} + C_1 } }}{{\left( {1 + r^2 } \right)^{\frac{1}{{M^4
\lambda ^2 }}} }}\,.
\end{equation}
Taking into consideration the delta function term, we obtain the solution
\begin{equation}
\psi \left( r \right) = C_2 \frac{{r^{\frac{2}{{M^4 \lambda ^2 }} + C_1 } }}{{\left( {1 + r^2 } \right)^{\frac{1}{{M^4
\lambda ^2 }}} }}e^{ - 4\frac{{\sigma r _0 }}{{M^4 }}\sqrt {\psi \left( {r _0 } \right)} \Theta \left( {r - r _0 } \right)
\ln \left( {\frac{r}{{r _0 }}} \right)} \,,
\end{equation}
which obviously reduces to the smooth case when we take the limit $r_0 \to 0$.
We note that the value $\psi(r_0)$, which enters in the exponent is just a constant factor.
This factor enters in (\ref{tuning1}) and, thus, the integration constant $C_2$ prevents any fine-tuning.
We also see that the transverse space retains a non-vanishing curvature inside the brane core.
There are also, in general,  conical singularities at $r=0$ and $r \to \infty$. However, we can
eliminate the singularity at the origin, by imposing that $C_1  =  - \frac{2}{{M^4 \lambda ^2 }}$. In
this case the 2-D metric is regular at $r=0$ and we only have a singularity at infinity, which signifies a
non-compact geometry. We can also check the existence of a deficit angle at infinity and demonstrate that no
singularity occurs as we cross the boundary of the resolved brane. To simplify our discussion, we define
 $b = 4\sigma r _0 M^{-4}\sqrt {\psi \left( r _0  \right)} $ and  $c = M^{-4} \lambda ^{-2} $. Expanding the conformal
factor around $r=r_0$, we obtain
\begin{equation}
\psi \left( r \right) \sim \frac{{C_2 }}{{\left( {1 + r _0 ^2 } \right)^c }}\,,
\end{equation}
so that the two radial coordinates ${ \rho }$ and $r$ are proportional near the brane and no deficit angle is involved.
However, for $r \to \infty$ we get
\begin{equation}
\psi \left( r \right) \sim r^{ - 2c} \left( {\frac{{r _0 }}{r}} \right)^b
\end{equation}
and the coordinate transformation yields a transverse space metric of the form
\begin{equation}
ds_2 ^2  = d \rho ^2  + k^2 \rho ^2 d\varphi ^2\,,
\end{equation}
where
\begin{equation}
k\,=\,1-c-\frac{b}{2}\,.
\end{equation}
The associated deficit angle is
\begin{equation}
\delta  = 2\pi \left( {1 - k} \right) = 2\pi \left( {c + \frac{b}{2}} \right)\,.
\label{deficit1}
\end{equation}
The combination $c+\frac{b}{2}$, which enters the expression for the deficit angle,
 is indeed the Euler number of the transverse space, as we can easily verify
\[
\chi  = \frac{1}{{4\pi }}\int {dr\,r\,\int\,d\varphi \,\psi \left( r \right)\,R^{\left( 2 \right)} }
 = \frac{1}{2}\int {dr\,r\,\left( {\frac{4}{{M^4 \lambda ^2 }}\frac{1}{{\left( {1 + r^2 } \right)^2 }}\, +
 \,\frac{{4\sigma }}{{M^4 }}\sqrt {\psi \left( {r _0 } \right)} \delta \left( {r - r _0 } \right)} \right)}
\]
\begin{equation}
 = \frac{1}{{M^4 \lambda ^2 }} + \frac{{2\sigma r _0 \sqrt {\psi \left( {r _0 } \right)} }}{{M^4 }}\, = \,c + \frac{b}{2}=\frac{\delta}{2\pi}\,.
\label{euler}
\end{equation}
In order for the space to have a finite volume, we must have
\begin{equation}
\frac{1}{{M^4 \lambda ^2 }} + \frac{{2\sigma r _0 }}{{M^4 }}\sqrt {\psi \left( {r _0 } \right)}  > 1\,,
\label{constraint1}
\end{equation}
which, given (\ref{euler}), is equivalent to $\chi>1$. The corresponding finite volume of the 2-D space turns out to be
\begin{equation}
V_2  = \pi C_2 \frac{{\left( {1 + r _0 ^2 } \right)^{1 - c} }}{{c - 1}} - i^{ - b - 2c} \pi C_2 r _0 ^b
B\left( { - \frac{1}{{r _0 ^2 }},c + \frac{b}{2} - 1,1 - c} \right)\,.
\end{equation}
$B$ is the incomplete Beta function. As a result of the positivity of parameters and the constraint (\ref{constraint1}),
 the volume is real and positive. Given the finite volume of the transverse space~\cite{Rub}-\cite{Keh}, this
 setup will exhibit a four-dimensional gravitational behavior at low energies compared to
 the compactification scale, mediated by a graviton zero mode on the brane and followed by a KK
 tower at higher energy scales. Apparently, the space remains non-compact for the entire range
 of parameter values for which the relation
 $1 < \frac{1}{{M^4 \lambda ^2 }} + \frac{{2\sigma r _0 }}{{M^4 }}\sqrt {\psi \left( {r _0 } \right)}  < 2$
  is satisfied. Eventually, for appropriate values, the Euler number of the transverse space
  reaches the value $\chi=2$ and the space compactifies into a sphere.

We could also place two branes in our setup, that would ensure a
compact transverse geometry from the beginning. To include the
second brane of tension $\sigma'$, situated at some position
$\rho_1>\rho_0$ away from the origin, we add to the action the
term
\begin{equation}
\int {d^5 xd\rho \delta \left( {\rho  - \rho _1 } \right)\sqrt { - \gamma }
 \left( {\sigma ' +
 \frac{1}{2}\gamma ^{\alpha \beta } \partial _\alpha  \bar \Sigma \partial _\beta  \bar \Sigma } \right)}\,,
\end{equation}
where the induced metric $\gamma$ is to be evaluated now at
$\rho=\rho_1$. With this addition, both branes remain flat as long
as we impose the condition
\begin{equation}
\sigma ' = \frac{\bar{q}^2}{2} g^{\phi \phi }\,,
\label{tuning2}
\end{equation}
relating the brane tension $\sigma'$ and the charge of the axion
$\bar q$ on the second brane. The metric component $g^{\phi \phi}$
is evaluated at $\rho=\rho_1$. Equation (\ref{ricci2}) now becomes
\begin{equation}
R^{\left( 2 \right)}  = \frac{4}{{M^4 \lambda ^2 }}\frac{1}{{\psi
\left( r \right)\left( {1 + r^2 } \right)^2 }} + \frac{{4\sigma
}}{{M^4 }}\frac{1}{{\sqrt {\psi \left( r \right)} }}\delta \left(
{r - r _0 } \right) + \frac{{4\sigma '}}{{M^4 }}\frac{1}{{\sqrt
{\psi \left( r \right)} }}\delta \left( {r - r _1 } \right)
\end{equation}
and the corresponding solution for the conformal factor reads
\begin{equation}
\psi \left( r \right) = C_2 \frac{{r^{2c + C_1 } }}{{\left( {1 + r^2 } \right)^c }}e^{ - b \Theta \left( {r - r _0 } \right)
\ln \frac{r}{{r _0 }}\ - b_1 \Theta \left( {r - r _1 } \right)\ln \frac{r}{{r _1 }}}\,,
 \end{equation}
where, in addition, we have defined $b_1  = 4M^{-4}\sigma 'r _1
\sqrt {\psi \left( r _1  \right)} $. Again, the metric is regular
at the origin if we set $C_1=-2c$. No deficit angle is encountered
around $r=0$ or as we cross each of the two branes. The total
deficit angle of the space is deduced by checking the metric at
infinity,
\begin{equation}
\psi \left( r \right) \sim r _0 ^b r _1 ^{b_1 } r^{2 - 2c - b -
b_1 }
\end{equation}
from which  the corresponding deficit angle
\begin{equation}
\delta  = 2\pi \left( {1 - k} \right) = 2\pi \left( {c + \frac{{b
+ b_1 }}{2}} \right) \label{deficit2}
\end{equation}
may immediately be obtained. To ensure that the space has the
topology of a sphere, we must impose the condition
\begin{equation}
\chi  = c + \frac{{b + b_1 }}{2} = 2\,,
\label{deficit3}
\end{equation}
which relates the tensions and positions of the two branes.

Let us take a moment here to discuss the way in which the
cosmological constant problem is addressed in the context of our
model. As we already pointed out, the effective 3-brane appears
Ricci-flat, with its tension inducing a deficit angle in the
non-compact transverse space. The relations connecting the various
physical constants in the case of the single brane setup are
(\ref{tuning1}) and (\ref{deficit1}). The pitfall of unwanted
fine-tunings may originate from these equations. To see if this is
the case, we imagine a situation where a flat brane solution has
been found for a specific brane tension $\sigma$. We then change
the value of the tension and check whether it induces a shift in
the deficit angle, which is unobservable, while leaving other
physical constants of the model unchanged. We see that such a
change may affect through (\ref{tuning1}) the value of the axion
field charge $q$. However, as we previously stressed, the metric
component $g^{\varphi \varphi}$ carries the undetermined
integration constant $C_2$, which in turn enters the constant
$\psi(r_0)$ and consequently $b$. Thus, changing the brane tension
leaves the axion charge unaltered and only affects the deficit
angle through (\ref{deficit1}), which involves $b$. In this way,
fine-tuning of physical constants in this setup is avoided. The
new brane tension could also lead to an additional change in the
radius of the 4-brane.

As it was also discussed in \cite{Kehagias:2004fb}, having a
non-compact transverse space is crucial. In the case of the
resolved double brane setup, however, there is still room for a
solution which doesn't require more than a mild fine-tuning,
despite the fact that the transverse space has necessarily the
topology of a sphere. This is due to relation (\ref{deficit3}),
which fixes the value of the previously undetermined constant
$C_2$. Taking the ratio of (\ref{tuning1}) and (\ref{tuning2}), we
see that altering the value of either $\sigma$ or $\sigma '$ will
induce a change in the ratio of the corresponding axion fields.
This can be avoided if the change in brane tension is compensated
by a change in the ratio of the metric components $g^{\varphi
\varphi}$ at $r_0$ and $r_1$. Since both carry the same overall
constant $C_2$, which cancels, the only way to satisfy this
requirement is for the brane radii to shift. Thus, a fine-tuning
between the two brane radii must be imposed to have flat branes
for arbitrary varying tensions. The four-dimensional cosmological
constant is again affecting the value of the deficit angle through
(\ref{deficit2}).

\section{Gauss-Codazzi Formulation}
An alternative way of investigating  the problem explored above is
by using the   Gauss-Codazzi formalism discussed in
\cite{Shiromizu:1999wj}. The
equations on a codimension-1 brane, when
  the bulk space is  $n$-dimensional turn out to be\footnote{While this paper was in
  preparation, a similar treatment of the brane equations was presented in \cite{Kanno:2007wj}, although in a
  cosmological context}
\[
{}^{(n - 1)}G_{\mu \nu }  = \frac{{n - 3}}{{n - 2}}\kappa_{n}^2
\left( {T_{\rho \sigma } q^\rho  _\mu  q^\sigma_\nu   + T_{\rho
\sigma } n^\rho  n^\sigma  q_{\mu \nu }  - \frac{1}{{n -
1}}Tq_{\mu \nu } } \right)
\]
\begin{equation}
 + KK_{\mu \nu }  - K_\mu  ^\sigma  K_{\nu \sigma }  - \frac{1}{2}q_{\mu \nu }
 \left( {K^2  - K^{\alpha \beta } K_{\alpha \beta } } \right) - E_{\mu \nu }\,,
 \label{geneq1}
\end{equation}
where $E_{\mu \nu}$ is the projection of the Weyl tensor of the bulk,
\begin{equation}
E_{\mu \nu }  = {}^{(n)}C^\alpha  _{\beta \rho \sigma } n_\alpha  n^\rho  q_\mu  ^\beta  q_\mu  ^\sigma
\end{equation}
and $K_{\mu \nu}$ is the extrinsic curvature tensor of the brane. The constant $\kappa_n$ is related to the
$n$-dimensional Planck mass of the theory. We will assume Gauss-normal coordinates, with the coordinate normal to the
brane denoted by $\rho$ and an energy-momentum tensor of the form
\begin{equation}
T_{\mu \nu }  =  - \Lambda g_{\mu \nu }  + T^{\left( B \right)} _{\mu \nu }  + S_{\mu \nu } \delta \left( \rho  \right)\,,
\end{equation}
where $T^{\left( B \right)} _{MN} $ is the bulk energy-momentum
tensor, not including the bulk cosmological constant. $S_{\mu\nu}$
is the energy-momentum content of the brane and it is further
decomposed in the our case according to
\begin{equation}
S_{\mu \nu }  =  - \sigma q_{\mu \nu }  + \tilde \tau _{\mu \nu }  + \bar \tau _{\mu \nu } \,.
\end{equation}
The first term represents brane tension, the second is the axion energy-momentum tensor and the
third accounts for additional matter content on the brane. This way we can also treat cases
with an arbitrary brane content in addition to the axion field which is only used to eliminate
 the azimuthal brane pressure and is effectively hidden. We proceed by further assuming a $Z_2$ symmetry in the $\rho$ direction.
 From Israel's junction conditions, this constraint fixes uniquely the extrinsic curvature of the brane in terms
 of its matter content, with the resulting expression being
\begin{equation}
K_{\mu \nu }  =  - \frac{{\kappa _n ^2 }}{2}\left( {S_{\mu \nu }  - \frac{1}{{n - 2}}q_{\mu \nu } S} \right)\,.
\end{equation}
Substituting this expression into (\ref{geneq1}) and using the
above mentioned energy-momentum tensors yields the effective
Einstein equation's on the brane
\begin{eqnarray}
{}^{(n - 1)}G_{\mu \nu } & =&  - \Lambda _{\left( {n - 1} \right)}
q_{\mu \nu }+ 8\pi \bar G_{\left( {n - 1} \right)} \bar T_{\mu \nu
} \nonumber \\
&&
   + 8\pi G_{\left( {n - 1} \right)} \left( {\tilde \tau _{\mu \nu }+ \bar \tau _{\mu \nu }
   }\right) + \kappa _n ^4 \left( {\tilde \pi _{\mu \nu }  + \bar \pi _{\mu \nu }  + \kappa _{\mu \nu } } \right) - E_{\mu \nu
   }\, .
\label{Gauss}
\end{eqnarray}
Quantities with tilde refer to the axion field contributions,
while barred quantities come from brane matter, the exception
being the second term which comes from the bulk content. The
definitions we use are
\begin{equation}
\Lambda _{\left( {n - 1} \right)}  = \left(\frac{{n - 3}}{{n - 1}}\right)\kappa _n ^2
 \left( {\Lambda  + \frac{{n - 1}}{{8\left( {n - 2} \right)}}\kappa _n ^2 \sigma ^2 } \right)\,,
\end{equation}
\begin{equation}
G_{\left( {n - 1} \right)}  = \left(\frac{{n - 3}}{{n - 2}}\right)\frac{{\kappa _n ^4 \sigma }}{{32\pi }}\,,
\end{equation}
\begin{equation}
\bar G_{\left( {n - 1} \right)}  = \left(\frac{{n - 3}}{{n - 2}}\right)\frac{{\kappa _n ^2 }}{{8\pi }}\,,
\end{equation}
\begin{equation}
\bar T_{\mu \nu }  = T^{\left( B \right)} _{\rho \sigma } q^\rho  _\mu  q^\sigma  _\nu   +
 T^{\left( B \right)} _{\rho \sigma } n^\rho  n^\sigma  q_{\mu \nu }
 - \frac{1}{{n - 1}}T^{\left( B \right)} q_{\mu \nu }\,,
\end{equation}
\begin{equation}
\pi _{\mu \nu }  =  - \frac{1}{4}\tau _\mu  ^\rho  \tau _{\nu \rho }  +
 \frac{1}{{4\left( {n - 2} \right)}}\tau \tau _{\mu \nu }  +
  \frac{1}{8}\tau ^{\alpha \beta } \tau _{\alpha \beta }  - \frac{1}{{8\left( {n - 2} \right)}}\tau ^2 q_{\mu \nu }\,,
\end{equation}
\begin{equation}
\kappa _{\mu \nu }  =  - \frac{1}{2}\tilde \tau _{(\mu } ^\rho  \bar \tau _{\nu )\rho }  +
\frac{1}{{4\left( {n - 2} \right)}}\left( {\tilde \tau \bar \tau _{\mu \nu }  +
 \bar \tau \tilde \tau _{\mu \nu } } \right) +
 \frac{1}{4}\tilde \tau ^{\alpha \beta } \bar \tau _{\alpha \beta } q_{\mu \nu }
  - \frac{1}{{4\left( {n - 2} \right)}}\tilde \tau \bar \tau q_{\mu \nu }\,.
\end{equation}
As a cross check of the formulas above, we note that by setting
$n=5$ and $\bar T_{\mu \nu }  = \tilde \tau _{\mu \nu } = 0$, we
recover the same equation derived in~\cite{Shiromizu:1999wj}.
Equation (\ref{Gauss}) is supplemented by Codazzi's equation
\begin{equation}
D_\nu  K_\mu  ^\nu   - D_\mu  K = \kappa _n ^2 T_{\rho \sigma } n^\rho  n^\sigma  q_\mu  ^\rho \,.
\end{equation}
Using (\ref{Gauss}), we can investigate the evolution on the
hypersurface of dimensionality $n-1$. Using (\ref{Gauss}), we can
investigate the evolution an $(n-1)$-dimsensional hypersurface by
applying the formalism described above.  We first consider the
model presented in \cite{Kaloper:2007my}, where we have an
effectively codimension-2 brane situated in a 6D bulk space.
 There is a 4-brane  and the axion field, without any additional matter in the bulk or the 4-brane and we
 also assume a $Z_2$ symmetry. The resulting space is flat everywhere, except from the position of the brane,
 where the tension induces a deficit angle in the bulk. Interestingly, it turns out
  that the imposed condition (\ref{tuning1}), relating the axion charge and the brane tension forces all terms to be
  quadratic in $\sigma$, so that the cosmological constant and ${\tilde \tau _{\mu \nu } }$ terms cancel against
  the $\tilde \pi _{\mu \nu }$ term. The equation of motion (\ref{Gauss}) reduces in this case to Einstein's
  equations in five dimensions, without any matter content,
\begin{equation}
^{\left( 5 \right)} G_{\mu \nu }  = 0\,.
\label{Einstein}
\end{equation}
We must stress here that $E_{\mu\nu}$ is evaluated near the brane and not on it, so in the
above mentioned space, which is flat outside the brane we get $E_{\mu\nu}=0$. By looking at
this equation and taking into consideration the fact that the angular coordinate is compact,
one is prematurely lead to think of this model as a regular Kaluza-Klein theory, with a zero mode
graviton and a tower of massive modes. This is however misleading, since (\ref{Einstein}) carries
no information of the fact that a dimensional reduction has already been performed on a non-compact dimension.
In the presence of matter the resulting equation is
\begin{equation}
^{\left( 5 \right)} G_{\mu \nu }  = 8\pi G_5 \bar \tau _{\mu \nu }  +
 \kappa _6 ^4 \bar \pi _{\mu \nu }  + \kappa _6 ^4 \kappa _{\mu \nu }  - E_{\mu \nu } \,.
\end{equation}
Notice that in this case the empty space solution no longer holds, so the bulk space will be curved in general and we can
no longer neglect the $E_{\mu\nu}$ term, which depends on the bulk curvature.

We now turn to the model we considered earlier, with the sigma
model fields in the bulk. By inspecting the energy-momentum tensor
for these fields, we see that the $T^{\left( B \right)} _{mn} $
components are zero in the coordinate system (\ref{coords}).
Notice that these coordinates are not Gauss-normal, but are
related to them through a conformal transformation. Since this
transformation doesn't mix $T^{\left( B \right)} _{mn} $ and
$T^{\left( B \right)} _{\mu \nu } $,  the former will also be zero
when we turn back to the Gauss-normal coordinates used to derive
equation (\ref{Gauss}). On the other hand, this transformation
will not affect the later components, which will be the same in
both cases. Assuming no additional bulk or brane matter besides
the axion and the $\phi^a$, the equations on the brane yield
\begin{equation}
^{\left( 5 \right)} G_{\alpha \beta }  =
\frac{{3\kappa _6 ^2 }}{4}\left( {\frac{{16}}{{5\lambda ^2 \left( {1 + r^2 } \right)^2\psi }}q_{\alpha \beta }
 - \frac{4}{{\lambda ^2 \left( {1 + r^2 } \right)^2\psi }}n_{\mu \nu } \delta ^\mu  _\alpha  \delta ^\nu  _\beta  } \right)
 - E_{\alpha \beta }\,.
\end{equation}
Here, $\alpha,\beta$ denote coordinates on the 4-brane, while
$\mu,\nu$ are coordinates on the 3-brane. The components of the
bulk energy-momentum tensor are taken as the limiting values near
the brane position. From this expression we immediately deduce
that ${}^{\left( 5 \right)}R = 0$, which means that the 4-brane
appears Ricci-scalar-flat, but not Ricci-flat. In fact, the fields
$\phi^a$ act on the brane as a form of perfect fluid with
anisotropic pressure, since we see that the fifth diagonal element
of the energy-momentum tensor is different. However, we still have
contributions from the Weyl tensor of the bulk space, which for
the metric (\ref{coords}) takes the form
\begin{equation}
E_{\alpha \beta }  =
\frac{{3\kappa _6 ^2 }}{{5\lambda ^2 }}\frac{1}{{\psi \left( {1 + r^2 } \right)^2 }}\,Diag\left(1, - 1, - 1, - 1,\,4\psi r^2
\right)\,.
\end{equation}
Once this is taken into account, we interestingly find that the
bulk field contribution  is cancelled by the projected Weyl tensor
and  the 4-brane has
\begin{equation}
^{\left( 5 \right)} R_{\mu \nu }  = 0 \, ,
\end{equation}
and thus, it is Ricci-flat in five dimensions.

Inclusion of matter on the brane is also straightforward in this
setup. In fact, the only additional contribution compared to the
previous model in empty bulk space comes from the sigma model
fields. The resulting equation is
\begin{eqnarray}
^{\left( 5 \right)} G_{\alpha \beta } & = & \frac{{3\kappa _6 ^2
}}{4}\left( {\frac{{16}}{{5\lambda ^2 \left( {1 + r^2 }
\right)\psi }}q_{\alpha \beta }
 - \frac{4}{{\lambda ^2 \left( {1 + r^2 } \right)\psi }}n_{\mu \nu } \delta ^\mu  _\alpha  \delta ^\nu  _\beta  }
 \right)\nonumber \\&&
  + 8\pi G_5 \bar \tau _{\alpha \beta }  + \kappa _6 ^4 \bar \pi _{\alpha \beta }  +
   \kappa _6 ^4 \kappa _{\alpha \beta }  - E_{\alpha \beta }\,.
\end{eqnarray}
To first order, we expect the first and last term to cancel as
before. In addition, we will have residual contributions in
$E_{\alpha \beta}$ due to the brane content.

\section{Conclusions}
We presented a model where
 effective codimension-2 branes are embedded in a 6-D bulk
 space endowed with an $O(3)$ sigma model. The introduction
 of resolved branes in such a background provides a more realistic approach compared
 to purely codimension-2 branes, which are known to be plagued by technical problems \cite{Cline:2003ak}.
 We presented solutions in the case of a single brane and non-compact
  transverse geometry and for a double brane setup in a sphere-compactified space.
  It turns out that the single brane setup can account for a flat 3-brane without requiring any
  fine-tunings of the physical parameters of the model. In the double brane scenario,
   a mild fine-tuning between the brane radii is needed. The Einstein equations on the effective codimension-2
   brane for arbitrary dimensionality were also derived and applied in our model
   to study the dynamics on the 4-brane. The model seems to admit a straightforward
   interpretation after KK reduction of the compact dimension of the brane. A more detailed
    future treatment of perturbations in this setup will help clarify further aspects of 4-D
    gravity and its possible modifications.

\vskip.3in

{\textbf{ Acknowledgments.}} This work was supported by the European
Research and Training Network MRTPN-CT-2006 035863-1
(UniverseNet). C. B. acknowledges also an
 {\textit{Onassis Foundation}} fellowship. Finally, K. T. acknowledges the hospitality of the CERN Theory Group.

\end{document}